\documentclass{elsart}
\usepackage{graphicx}

\begin{document}
 
\begin{frontmatter}   
 
\title{Loss of  superfluidity  in
a
Bose-Einstein condensate on an optical lattice via a novel classical phase
transition}
 
\author{Sadhan K. Adhikari}
\address{Instituto de F\'{\i}sica Te\'orica, Universidade Estadual
Paulista, \\ 01.405-900 S\~ao Paulo, S\~ao Paulo, Brazil}

\date{\today}
\maketitle

\begin{abstract}

We predict the loss of superfluidity in a Bose-Einstein condensate (BEC) 
trapped in a combined optical and axially-symmetric harmonic potentials
during a resonant collective excitation initiated by a periodic modulation
of the atomic scattering length $a$, when the modulation frequency equals
twice the radial trapping frequency or multiples thereof.  This classical
dynamical transition is marked by a loss of superfluidity in the BEC and
a subsequent destruction of the interference pattern upon free expansion. 
Suggestion for future experiment is made.

\end{abstract}
   
 
\end{frontmatter}

The detailed study of quantum phase
effects on a macroscopic scale such as interference of matter
waves \cite{kett1} has been possible after 
the experimental loading of a cigar-shaped Bose-Einstein condensate 
(BEC) in a combined axially-symmetric harmonic plus a standing-wave 
optical lattice potential
traps in   both one \cite{1,2} and three \cite{greiner} dimensions. There have been
several theoretical studies on different aspects of a BEC in a one-
\cite{th,adhi1} as well as three-dimensional \cite{adhi} optical-lattice
potentials. The phase coherence among different optical-lattice sites of a
trapped BEC on an optical lattice has been established in recent
experiments \cite{1,2,greiner,cata,catax} through the formation of
distinct interference pattern when the traps are removed. This long-range
phase coherence in the condensate along the entire optical lattice is a
sign of communication among various sites which is necessary for
developing superfluidity in the condensate.

The phase-coherent BEC on the optical lattice is a superfluid
\cite{greiner,stoof} as the atoms in it move freely from one optical site
to
another by quantum tunneling through the high optical potential
barriers. It has been
demonstrated in a recent experiment by Greiner {\it et al.} \cite{greiner}
that, as the optical potential traps are made much too higher, the quantum
tunneling of atoms from one optical site to another is stopped resulting
in a loss of superfluidity in the BEC. Consequently,
no interference pattern is formed upon free expansion of such a BEC which
is termed a Mott insulator state \cite{greiner,stoof}, in which  
an individual atom is  attached to a fixed optical site
and its free mobility to a nearby site by tunneling is stopped as in an
insulator. This phenomenon represents a superfluid-insulator quantum
phase transition 
and cannot be properly accounted for in a mean-field model based on the
Gross-Pitaevskii (GP) equation \cite{8} where these quantum effects are
mostly lost.

Following a suggestion by Smerzi {\it et al.} \cite{sm}, Cataliotti {\it
et al.} \cite{cata2,cata3} have demonstrated in a novel experiment the
loss of superfluidity in a BEC trapped in a one-dimensional
optical-lattice and harmonic potentials when the center of the harmonic
potential is suddenly displaced along the optical lattice through a
distance larger than a critical value.  Then a modulational instability
takes place in the BEC. Consequently, it cannot reorganize itself quickly
enough and the phase coherence and superfluidity of the BEC are lost.  
The loss of superfluidity is manifested in the destruction of the
interference pattern upon free expansion.  Distinct from the quantum phase
transition observed by Greiner {\it et al.} \cite{greiner}, this
modulational instability responsible for the superfluid-insulator
transition is classical dynamical in nature \cite{sm,cata2,cata4}. This
process is also different from the Landau dissipation mechanism
\cite{sm,cata4}, occurring when the fluid velocity is greater than local
speed of sound. The present dynamical phase transition can be well
described by the mean-field GP equation \cite{adhi1,sm,cata4}. When Landau
instability occurs, the system lowers energy by emitting phonons
\cite{cata4}. The GP equation can not simulate energy dissipation and
hence the Landau dissipation mechanism.

The above modulational instability is not the unique dynamical classical
process leading to a superfluid-insulator transition. Many other classical
processes leading to a rapid movement or collective excitation in the
condensate may lead to such a transition. The movement should be rapid
enough so that the BEC cannot reorganize itself to evolve through orderly
phase coherent states.  Here we suggest that a collective resonant
excitation of the BEC  may also lead to
the destruction of superfluidity due to a classical superfluid-insulator
transition. There have been theoretical \cite{osc} and experimental
\cite{3} studies of collective excitation in the BEC in the absence of an
optical trap initiated by a modulation of the trapping frequency. The
study of such collective excitation in the presence of an optical trap has
just began \cite{3a}.

In the present study the collective excitation  is initiated near a
Feshbach resonance \cite{mit} by a
periodic modulation   of the repulsive atomic scattering length $a$ $ 
(>0)$
 via $a \to a +\bar A\sin(\Omega t)$ where
$t$ is time, $\bar A$ an amplitude, 
and $\Omega$ the frequency of modulation.  
Such modulation of the  scattering length  can be realized
experimentally near a Feshbach
resonance by manipulating an external background magnetic field. Although,
the background magnetic field and the scattering length are nonlinearly
related in general, for small modulations $\bar A$ $(<a)$ an approximate
linear
relation between the background magnetic field and the scattering length
may  hold which might make  the implementation of the above
modulation experimentally possible.
When  $\Omega= 2\omega$ or  multiples thereof, 
resonant collective oscillation can be generated  in the BEC,  where
$\omega$ is the 
radial trapping frequency \cite{adhi2}. This resonant oscillation 
destroys the superfluidity of the BEC provided that the condensate is
allowed to experience this oscillation for a certain interval of
time called hold time. 
We base the present study on the numerical solution of the
time-dependent mean-field
axially-symmetric GP equation \cite{8} in the presence
of a
combined 
harmonic and optical potential traps. This transition involving collective
excitation of the BEC and 
described by the
mean-field GP equation is classical, rather than quantum,  in nature.

The time-dependent BEC wave
function $\Psi({\bf r};\tau)$ at position ${\bf r}$ and time $\tau $
is described by the following  mean-field nonlinear GP equation
\cite{8}
\begin{eqnarray}\label{a} \left[- i\hbar\frac{\partial
}{\partial \tau}
-\frac{\hbar^2\nabla^2   }{2m}
+ V({\bf r})
+ gN|\Psi({\bf
r};\tau)|^2
 \right]\Psi({\bf r};\tau)=0,
\end{eqnarray}
where $m$
is
the mass and  $N$ the number of atoms in the
condensate,
 $g=4\pi \hbar^2 a/m $ the strength of interatomic interaction.  
In the presence of the combined
axially-symmetric and optical lattice traps 
     $  V({\bf
r}) =\frac{1}{2}m \omega ^2(r^2+\nu^2 z^2) +V_{\mbox{opt}}$ where
 $\omega$ is the angular frequency of the harmonic trap 
in the radial direction $r$,
$\nu \omega$ that in  the
axial direction $z$, with $\nu$ the aspect ratio, and $V_{\mbox{opt}}$ is
the optical lattice trap introduced later.  
The normalization condition  is
$ \int d{\bf r} |\Psi({\bf r};\tau)|^2 = 1. $

In the axially-symmetric configuration, the wave function
can be written as 
$\Psi({\bf r}, \tau)= \psi(r,z,\tau)$.
Now  transforming to
dimensionless variables $x =\sqrt 2 r/l$,  $y=\sqrt 2 z/l$,   $t=\tau \omega, $
$l\equiv \sqrt {\hbar/(m\omega)}$,
and
${ \varphi(x,y;t)} \equiv   x\sqrt{{l^3}/{\sqrt
8}}\psi(r,z;\tau),$  Eq. (\ref{a}) becomes \cite{11}
\begin{eqnarray}\label{d1}
&\biggr[&-i\frac{\partial
}{\partial t} -\frac{\partial^2}{\partial
x^2}+\frac{1}{x}\frac{\partial}{\partial x} -\frac{\partial^2}{\partial
y^2}
+\frac{1}{4}\left(x^2+\nu^2 y^2\right) \nonumber \\
&+&\frac{V_{\mbox{opt}}}{\hbar \omega} -{1\over x^2}  +                                                          
8\sqrt 2 \pi n\left|\frac {\varphi({x,y};t)}{x}\right|^2
 \biggr]\varphi({ x,y};t)=0, 
\end{eqnarray}
where nonlinearity
$ n =   N a /l$. In terms of the 
one-dimensional probability 
 $P(y,t) \equiv 2\pi\- \- \int_0 ^\infty 
dx |\varphi(x,y,t)|^2/x $, the normalization  
is given by $\int_{-\infty}^\infty dy P(y,t) = 1.$

We use the parameters of 
the  experiment of Cataliotti {\it et al.} \cite{cata}
with repulsive $^{87}$Rb atoms where  
the radial trap frequency was  $ \omega =
2\pi \times 92$ Hz. The
optical
potential created with the standing-wave laser field of wavelength 
$\lambda=795$ nm is given by $V_{\mbox{opt}}=V_0E_R\cos^2 (k_Lz)$,
with $E_R=\hbar^2k_L^2/(2m)$, $k_L=2\pi/\lambda$, and $V_0$ $ (<12)$ the 
strength. For the mass $m=1.441\times 10^{-25}$ kg of $^{87}$Rb the
harmonic
oscillator length $l=\sqrt {\hbar/(m\omega)} = 1.126$ $\mu$m and 
and the 
dimensionless time unit  $\omega ^{-1} =
1/(2\pi\times 92)$ s $ =1.73$ ms. In terms of the dimensionless laser wave
length $\lambda _0= \sqrt2\lambda/l \simeq 1$ and  the dimensionless 
 energy  $E_R/(\hbar \omega)= 4\pi^2/\lambda _0^2$,  $V_{\mbox{opt}}$ of 
Eq.  (\ref{d1}) is
$
{ V_{\mbox{opt}}}/{(\hbar \omega)}=V_0({4\pi^2}/{\lambda_0^2}) 
\cos^2 (
{2\pi}y/{\lambda_0}).
$

We solve  Eq.  (\ref{d1}) numerically  using a   
split-step time-iteration
method
with  the Crank-Nicholson discretization scheme described recently
\cite{11}.  
The time iteration is started with the known harmonic oscillator solution
of  Eq.  (\ref{d1}) with
 $n=0$: $\varphi(x,y) = [\nu
/(8\pi^3)  ]^{1/4}$
$xe^{-(x^2+\nu y ^2)/4}$ 
\cite{11}. 
The
nonlinearity $n$  as well as the optical lattice strength $V_0$ 
are  slowly increased by equal amounts in $10000n$ steps of 
time iteration until the desired nonlinearity and 
optical lattice
potential are  attained. Then, without changing any
parameters, the solution so obtained is iterated 50 000 times so that a
stable
solution  is obtained 
independent of the initial input
and time and space steps.

The one-dimensional pattern of  BEC  on the
optical
lattice  for a specific nonlinearity and the 
interference pattern upon its free expansion  have been
recently studied using the numerical solution of Eq.
 (\ref{d1}) \cite{adhi1}. Here we study the destruction of this
interference pattern after the application of a periodic modulation of 
the scattering length resulting in a similar modulation of nonlinearity
$n$ in
Eq. (\ref{d1}) via 
\begin{equation}\label{rep}
n  \to n +A\sin (\Omega t),
\end{equation}where $A$ is an amplitude.
 In the present
model
study we employ  nonlinearity $n=5$,  the axial trap parameter 
$\nu =0.5$, 
and   the optical lattice strength 
$V_0=6$ throughout. First we calculate  the ground-state wave
function in the combined harmonic  and optical lattice potentials.

 When
the
condensate is released from the combined trap, a matter-wave  
interference
pattern is
formed in a few milliseconds as described in Ref. \cite{adhi1}.  The atom
cloud released from one lattice
site expand, and overlap and interfere with atom clouds from neighboring
sites to form the robust interference pattern due to phase coherence.
The pattern consists of a central peak 
and two symmetrically spaced peaks, each containing about $10\%$ of 
total number of atoms,
moving in opposite directions \cite{cata,catax,adhi1,xxx}.

\begin{figure}[!ht]
 
\begin{center}

\includegraphics[width=.8\linewidth]{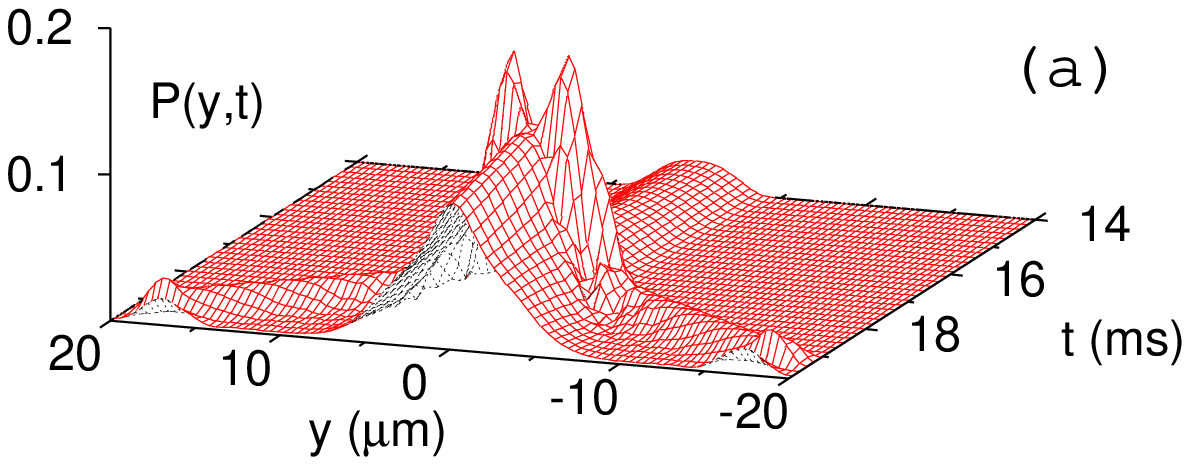}
\includegraphics[width=.8\linewidth]{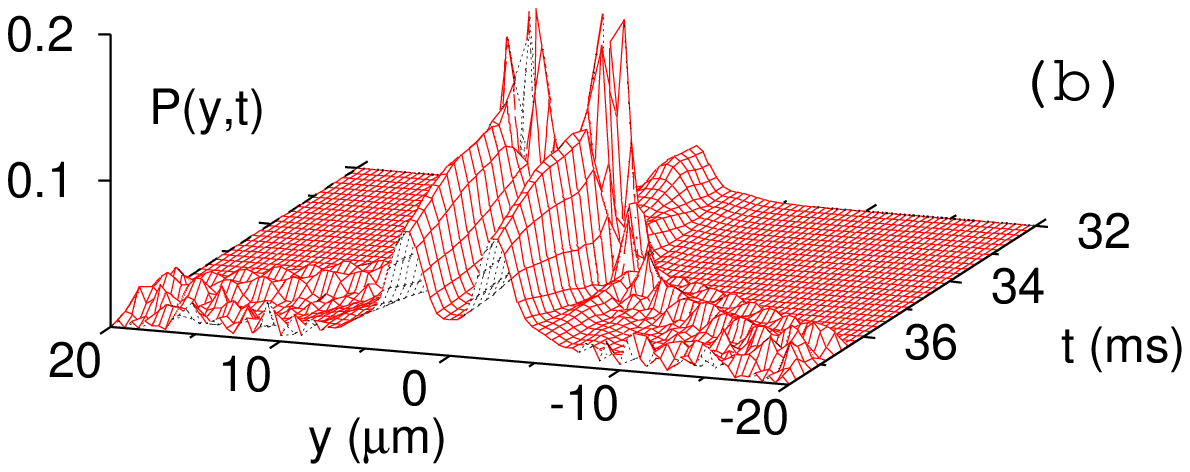}
\includegraphics[width=.8\linewidth]{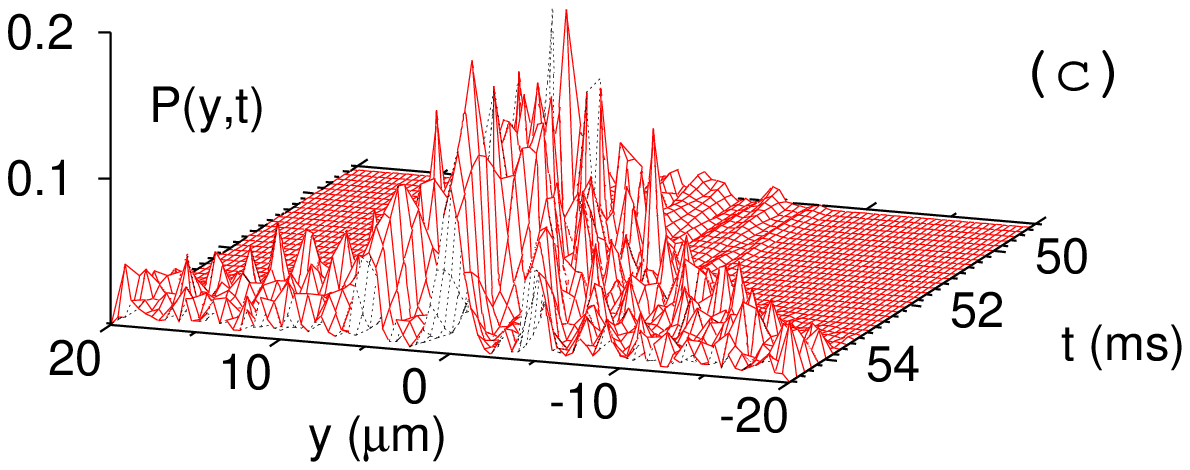}
\end{center}
 
\caption{One-dimensional probability  $P(y,t)$
vs. $y$ and $t$ for the BEC on optical lattice 
under the action of modulation (\ref{rep}) with  $n=5$, $\Omega =
2\omega$
and $A=3$ 
and upon the removal of the
combined traps after hold times (a) 17 ms, (b) 35 ms, and (c) 52 ms.  
} \end{figure}

\begin{figure}[!ht]
 
\begin{center}

\includegraphics[width=.8\linewidth]{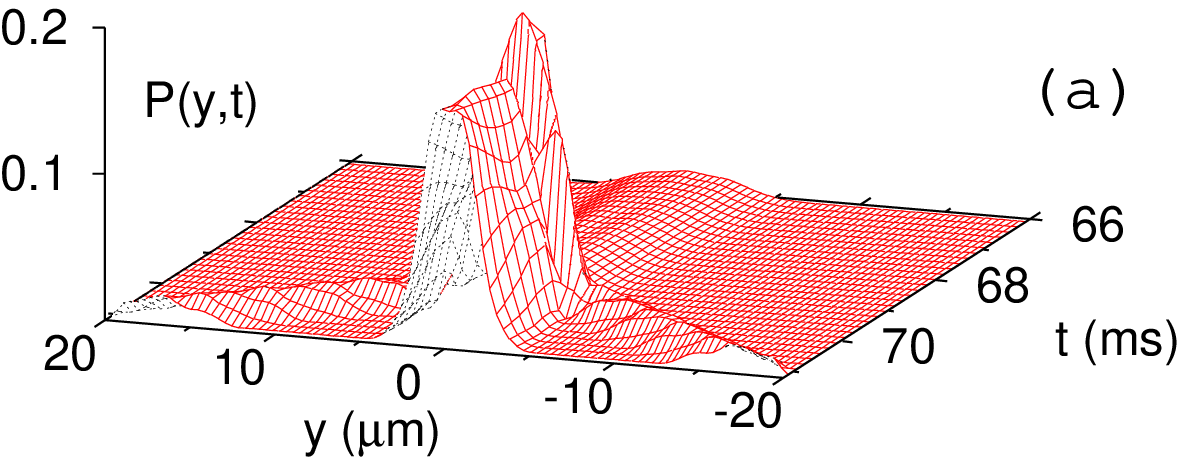}
\includegraphics[width=0.8\linewidth]{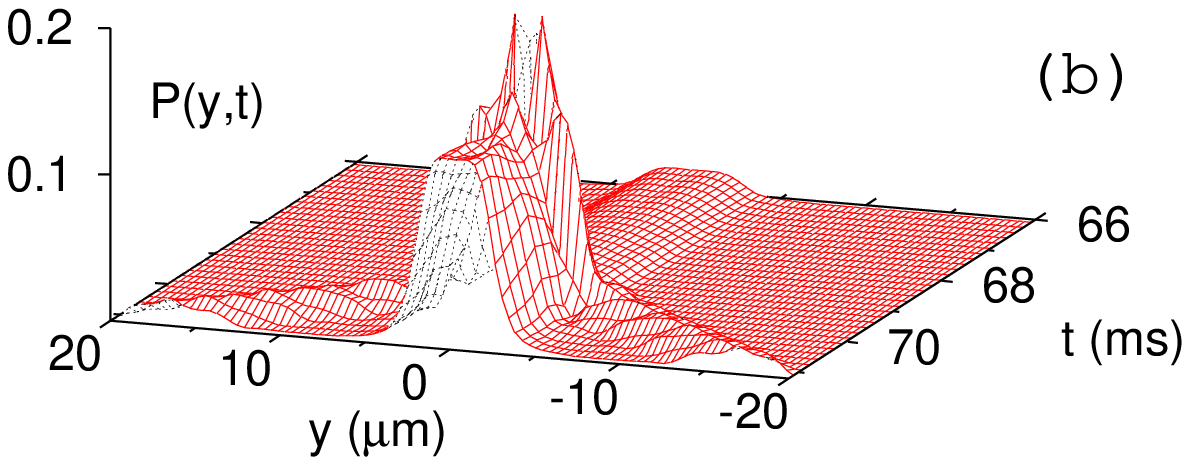}
\end{center}
 
\caption{One-dimensional probability  $P(y,t)$
vs. $y$ and $t$ for the BEC on optical lattice 
under the action of modulation (\ref{rep}) with (a)  $\Omega =\omega$
and (b)   $\Omega =3\omega$ 
with $n=5$ and   $A=3$ 
 upon the removal of the
combined traps after hold time 69 ms.
} \end{figure}

If we introduce the modulation of  nonlinearity (\ref{rep})  
after the formation of the BEC in the combined trap, the condensate will
be out of equilibrium and start to oscillate. As the height of the
potential  barriers in the optical lattice is much larger than the
energy of the system, the atoms in the condensate will move by tunneling
through the potential barriers. This fluctuating transfer of Rb atoms
across the potential barriers is due to Josephson effect in a neutral
quantum liquid \cite{cata}. The phase coherence between different optical
sites  of
the condensate may be destroyed during this rapid oscillation
with large amplitude and no matter-wave interference
pattern will be formed after the removal of the joint traps.

However, resonant oscillation arising from modulation (\ref{rep}) is not
the only classical dynamical mechanism for the destruction of
superfluidity. It can also be destroyed  via a periodic
modulation of the
radial trapping potential  \cite{adhi5} or by giving the BEC a large
displacement
along the optical trap \cite{catax}. For a weak modulation or small
displacement, the superfluidity
is
preserved independent of hold time in the modulated trap. For a stronger
modulation and longer hold time, there is destruction of superfluidity via
a classical dynamical process.

Now we explicitly study the destruction of superfluidity in the condensate
upon application of the modulation (\ref{rep}) leading to a resonant
oscillation. The loss of superfluidity only takes place if  the BEC is
allowed to experience the resonant oscillation  for a
certain interval of time (hold time).  The resonant oscillation is 
excited for  $\Omega=2\omega$ or multiples thereof. Strictly speaking,
this should happen when $\Omega$ equals the natural frequency of
oscillation of the system  or multiples thereof. This  natural frequency
is $2\omega$ for a noninteracting system with $n=0$. In the presence of
nonlinear mean-field interaction this frequency is expected to
change. This change is found to be  small in numerical
simulation in the present context of small
nonlinearity. However, some deviation is found to take place for 
large nonlinearity \cite{11}.

  In this calculation we take $n=5$ in Eq. (\ref{rep}) and
for
the effective nonlinearity after modulation to remain positive (repulsive 
condensate)  we restrict to
$A<5$. Negative values of nonlinearity corresponding to atomic attraction
may
lead to collapse and instability \cite{8} and will not be considered here.
We shall present results with $A=3$ in this study, although any
other $A$, which is not negligibly small, leads to similar result.  For
$\Omega=2\omega$ the BEC executes rapid oscillation exciting collective
resonant modes \cite{3,3a} resulting is a destruction of superfluidity.

For numerical simulation we allow the BEC to evolve on  a lattice with 
$x \le 20$ $\mu$m and 20 $\mu$m $\ge y\ge$ $-20$ $\mu$m after the
modulation (\ref{rep}) is applied 
and study
the system after different   hold times.
The one-dimensional
probability  $P(y,t)$ is  plotted in figures 1 (a), (b)   and (c)
for hold times 17 ms,  35 ms and  52 ms,  respectively. For 
hold
time  17 ms,  prominent interference
pattern is
formed upon free expansion.
In figure 1 (a)  
three separate pieces in the  interference pattern  
corresponding  to three distinct trails can be identified. 
The
interference 
pattern is slowly destroyed at  increased  hold times as  we can see in
figures 1 (b)  and (c). 
As the hold time  increases the maxima
of the interference pattern
mix up upon free expansion  and finally for the  hold time of 
52 ms  the interference pattern is
completely destroyed as we find in figure 1 (c).
As the BEC is allowed to evolve for a substantial  interval of time after
the application of the periodic modulation in the scattering length, a
dynamical instability of classical nature sets in which destroys the
superfluidity \cite{sm,cata2}.

The superfluidity  reappears rapidly as  
frequency $\Omega $ of the modulation (\ref{rep}) is changed  to a
nonresonant value. 
We demonstrate this for   $\Omega=\omega  $ and $3 \omega$  in the
following.
In figures 2 (a) and (b) we present the evolution of probability 
$P(y,t)$ after the application of the  modulation  for  $n=5$, $A= 3$ and 
$\Omega =\omega  $ and $3 \omega$. We see from figures 2 (a) and (b) that
in both cases the interference pattern is obtained  after a hold time of 
69 ms.

\begin{figure}[!ht]
 
\begin{center}

\includegraphics[width=.8\linewidth]{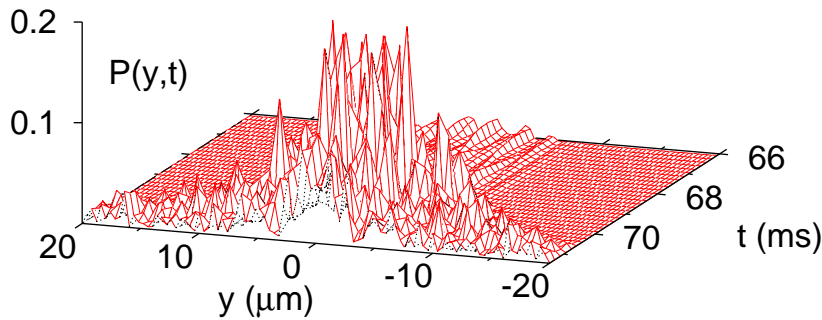}
\end{center}
 
\caption{One-dimensional probability  $P(y,t)$
vs. $y$ and $t$ for the BEC on optical lattice 
under the action of modulation (\ref{rep}) with   $\Omega =4\omega$
with $n=5$ and  $A=3$  upon the removal of the
combined traps after hold time 69 ms.
} \end{figure}

The resonant collective oscillation  also appears for higher multiples of 
$\Omega = 2\omega$ thus leading to a destruction of superfluidity. This
is illustrated in Fig. 3 where 
 we present the evolution of probability
$P(y,t)$ after the application of the  modulation  with $n=5$, $A= 3$ and
$\Omega =4\omega  $ for a hold time of 69 ms. We see that no interference
pattern is formed in this case upon release of the BEC from the 
joint traps.  In this case the superfluidity  is maintained for a hold
time of 52 ms whereas for $\Omega = 2\omega$ it was destroyed at 52 ms
 as can be seen in figure 1 (c).

In the absence of the optical trap, a cigar-shaped BEC  
can be excited to  collective resonant states for modulation (\ref{rep})
when $\Omega$ is an even multiple of $\omega$ or $\nu \omega$
\cite{adhi2}. The present study shows that in the presence of an optical
trap, only the former possibility leads to a loss of superfluidity, at
least for the small nonlinearity ($n=5$) considered here.
The possibility $\Omega=2\nu \omega$  does not
seem to lead to prominent collective resonant excitation and hence does
not easily lead to a loss of superfluidity as one can see from figure 2
(a) with $\nu = 0.5$. The reason for the absence of collective resonant
excitation in this case is not clear. It is possible that for large
nonlinearity $n$ and amplitude $A$ in (\ref{rep}) there is breakdown of
superfluidity in this case.

In conclusion, 
we have studied  the destruction of superfluidity 
in a cigar-shaped BEC loaded in a combined 
harmonic  and  optical lattice traps  upon the application of 
a periodic modulation of the scattering length leading to resonant
collective excitation when the frequency of modulation equals twice
the radial trapping frequency or multiples thereof.   In the absence of 
modulation,  the
formation of the interference pattern  upon the removal of the combined
traps clearly demonstrates the
phase coherence \cite{adhi,adhi1}. 
At these resonance frequencies  
the phase coherence is destroyed 
signaling a superfluid-insulator classical phase transition,
provided
that 
the BEC is kept in the modulated trap for a certain hold time.
Consequently,
after release from the combined trap no interference pattern is
formed. The superfluidity  in the BEC  is quickly restored when the
frequency of
modulation of the scattering length is changed to a nonresonant value
away from $\Omega = 2\omega$ or multiples. 
It is possible to study this novel superfluid-insulator classical phase
transition  experimentally and a comparison of those results  with
mean-field models will 
enhance our understanding  of matter wave BEC. After the recent
experiments
of Cataliotti {\it et al.} \cite{cata2} and M\"uller  {\it et al.}
\cite{xxx}
that study seems possible in a not too distant future.

\ack

The work was supported in part by the CNPq and FAPESP
of Brazil.


 \end{document}